\documentclass[aps,letter,prb,twocolumn]{revtex4-1}
	
	\usepackage{amssymb}
	\usepackage{amsmath}
	\usepackage{amsthm} 
	\usepackage{bm}
	\usepackage{enumerate}
	\usepackage{graphicx}
	\usepackage[caption=false]{subfig}
	\usepackage[version=3]{mhchem}
	\usepackage{dcolumn} 
	\usepackage[colorlinks,citecolor=blue,linkcolor=blue,urlcolor=blue]{hyperref}
	
	\newcommand{\avg}[1]{\left< #1 \right>} 
	\newcommand{\trace}{\mbox{Tr}} 
	
\begin{document}
	\title{Noncollinear magnetic ordering in the Shastry-Sutherland Kondo lattice model: Insulating regime and the role of Dzyaloshinskii-Moriya interaction}
	\author{Munir~Shahzad and Pinaki~Sengupta}
	\affiliation{School of Physical and Mathematical Sciences, Nanyang Technological University, 21 Nanyang Link, Singapore 637371}
	\date{\today}
	\begin{abstract}
				
	We investigate the necessary conditions for the emergence of complex, noncoplanar magnetic configurations in a Kondo lattice model with classical local moments on the geometrically frustrated Shastry-Sutherland lattice and their evolution in an external magnetic field. We demonstrate  that topologically nontrivial spin textures, including a new canted flux state, with nonzero scalar chirality arise dynamically from realistic short-range interactions. Our results establish that a finite Dzyaloshinskii-Moriya (DM) interaction is necessary for the emergence of these novel magnetic states when the system is at half filling, for which the ground state is insulating. We identify the minimal set of DM vectors that are necessary for the stabilization of chiral magnetic phases.  The noncoplanarity of such structures can be tuned continually by applying an external magnetic field. This is the first part in a series of two papers; in the following paper the effects of frustration, thermal fluctuations, and magnetic field on the emergence of novel noncollinear states at metallic filling of itinerant electrons are discussed. Our results are crucial in understanding the magnetic and electronic properties of the rare-earth tetraboride family of frustrated magnets with separate spin and charge degrees of freedom.
	
	\end{abstract}
	\maketitle
	
	\section{Introduction}\label{sec:intro}
	
The study of strongly interacting quantum many-body systems with independent spin and charge degrees of freedom on frustrated lattices has attracted heightened interest in the recent past. The interplay between geometric frustration and strong interaction 
between itinerant electrons and localized moments in these systems results in novel quantum phases and phenomena that are not observed in their nonfrustrated counterparts~\cite{martin-2008,barros-2014,ishizuka-2013ll,chern-2015,yasu-2010,
machida-2007,nakatsuji-2006,udagawa-2013}.
These competing interactions, together with crystal electric fields and coupling to the itinerant electrons, often stabilize
	noncoplanar ordering of these moments~\cite{martin-2008,chern-2010,akagi-2011,akagi-2012,venderbos-2012}. When an 
	electron moves through such background spin textures, it picks up a Berry phase which underlies several novel transport phenomena such as the topological (or geometric) Hall effect 
	and unconventional magnetoresistive properties~\cite{karplus-1954,ye-1999,niu-2010,taguchi-2001}. 
	The interest in these systems is driven by the desire both to understand the underlying 
	mechanism driving the novel phenomena and to control their emergence by external tuning 
	fields in order to harness their unique functionalities for practical applications. 
			
	In this paper, we study the Kondo lattice model (KLM) on the geometrically frustrated Shastry-Sutherland lattice (SSL) with classical spins where the standard (antiferromagnetic) Heisenberg interaction between the local moments is supplemented by an additional Dzyaloshinskii-Moriya (DM) interaction. The SSL is a paradigmatic geometry to study the effects of competing interactions in the presence of frustration~\cite{shastry-1981}. The Shastry-Sutherland Kondo lattice model (SS-KLM) has previously been studied with $S=1/2$ local moments~\cite{coleman-2010,bernhard-2011,pixley-2014,lei-2015}, where quantum fluctuations of the local moments play a crucial role in determining the character of the ground state. In the present study, we revisit this model, but with the local moments treated as classical spins. This is not simply of academic interest. There exists a complete family of rare-earth tetraborides ($R$\ce{B_4}, $R$=Tm, Er, Ho, Dy),
	quasi-two-dimensional metallic magnets in which the magnetic-moment-carrying rare-earth ions are arranged in a SSL in the layers. Due to strong spin-orbit coupling, the rare-earth ions in these compounds carry large magnetic moments and,
	consequently, can be treated as classical spins. They act as effective local fields that interact strongly with the electron spin~\cite{siemensmeyer-2008,keola-2015,sai-2016,suzuki-2010,matas-2010,suzuki-2009,iga-2007,shinji-2009}.
	In this  study our goal is to construct a minimal model where topologically nontrivial chiral 
	magnetic phases can be realized from physically relevant interactions and investigate their 
	evolution in an external magnetic field. In particular, we explore the role of different components 
	of the DM interaction in stabilizing different aspects of the local-moment configurations. 
	What are the minimal DM vectors required to stabilize a {\em tunable} noncoplanar spin configuration? How does an applied field affect the noncoplanarity of the spin configuration? Does the {\em nature} of the chiral spin state change in the presence of an external field? These are some of the questions we address in this work. Our results reveal that multiple noncoplanar spin arrangements (characterized by different values of the scalar spin chirality) with long-range magnetic order are stabilized over an extended range of parameters.
	Not surprisingly, we find that DM interactions play a crucial role in stabilizing chiral spin
	configurations. Furthermore, we are able to tune the noncoplanarity (equivalently, the topological character) of the spin textures, changing and suppressing the net chirality, by applying an external magnetic field. This is in contrast to most previous studies in which the noncoplanar textures of the local moments were imposed by extraneous factors (e.g., crystal electric field in pyrochlores) and as such cannot be changed easily. In an accompanying paper~\cite{shahzad-2017}, we follow this up by studying the role of thermal fluctuations, frustration, and magnetic field in stabilizing the noncollinear magnetic states and phase transitions at one-quarter and three-quarter filling of itinerant electrons, for which the ground state is metallic. Moreover, we found out that unlike the insulating case (discussed in this paper),  the noncollinear textures emerge for metallic densities without a DM interaction between the local moments. 
			
	\section{Model}\label{sec:model}
						
\begin{figure}[htb]
	\centering
	\includegraphics[clip,trim=0cm 0cm 0cm 0cm,width=\linewidth]{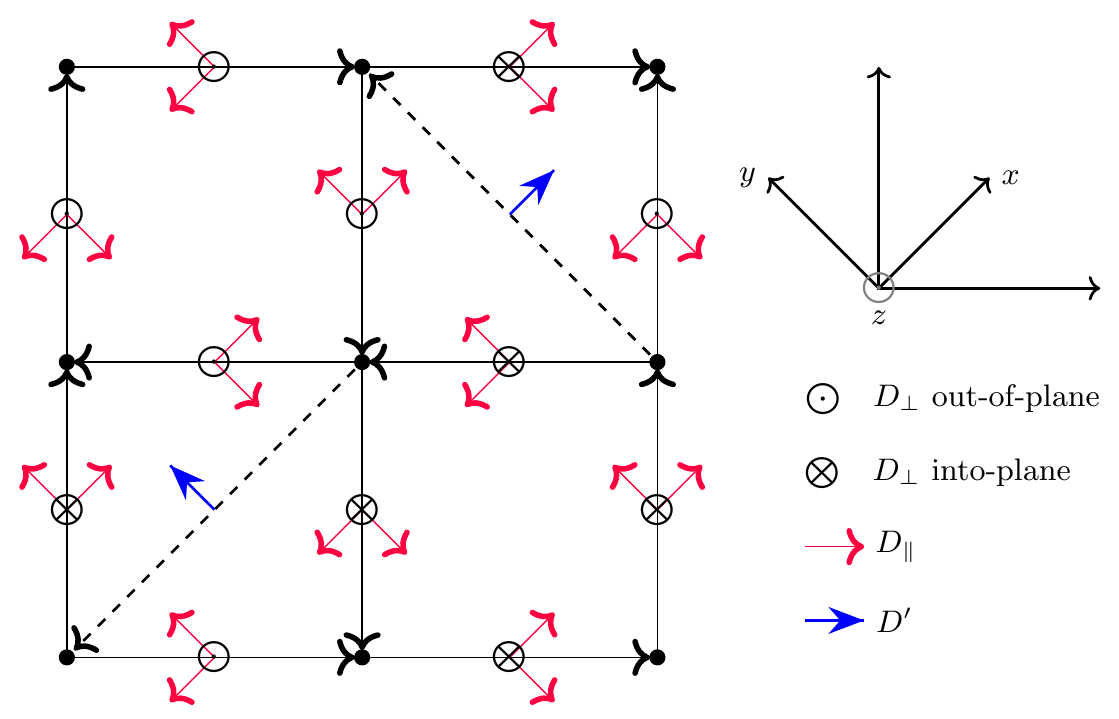}
	\caption{(Color online) DM interaction defined on the unit cell of SSL where the direction of the arrow from site $i$ to site $j$ indicates the direction of cross product $\mathbf{S}_i\times \mathbf{S}_j$. The red arrows represent the parallel components of $\mathbf{D}$, while $\bigodot$ and $\bigotimes$ represent the out-of-plane and into-plane components of $\mathbf{D}$. Blue arrows indicate the components of $\mathbf{D}'$ on the diagonal bonds. The directions of $x$, $y$, and $z$ axes are also indicated.}
	\label{fig:SSL-dm}
	\end{figure}

	The Hamiltonian describing the SS-KLM with additional DM interactions is given by
				
	\begin{equation}
	\label{equ:hamiltonian}
	\mathcal{\hat{H}}= {\mathcal{\hat{H}}_e} + {\mathcal{\hat{H}}_c},
	\end{equation}
	where $\mathcal{\hat{H}}_e$ represents the electronic Hamiltonian,
				
	\begin{equation}
	\label{equ:elec part}
	{\mathcal{\hat{H}}_e}= -\sum_{\avg{i,j},\sigma}t_{ij}(c_{i\sigma}^\dagger c_{j\sigma}+\mbox{H.c.})-J_K \sum_i \mathbf{S}_i\cdot \mathbf{s}_i.
	\end{equation}
	The first term is the kinetic energy of the itinerant electrons; $\avg{i,j}$ represents the Shastry-Sutherland bonds (viz., first neighbors along the principal axes and the
	second neighbors along the alternate diagonals), and $t_{ij}$ are the transfer integrals for these bonds. The second term is the on-site Kondo-like interaction between the spin of the itinerant electrons $\mathbf{s}_i$ and localized moments $\mathbf{S}_i$ . The conduction electron spin is defined as $\mathbf{s}_i= c_{i,
	\alpha}^\dagger\bm{\sigma}_{\alpha\beta}c_{i,\beta}$, where $\bm{\sigma}_{\alpha \beta}$ is the vector element of the usual Pauli matrices. As mentioned in the Introduction, we treat the localized spins as classical vectors with unit length ($\vert \mathbf{S}_i\vert = 1$). In this limit, the sign of $J_K$ (ferromagnetic or antiferromagnetic) is irrelevant  since eigenstates that correspond to different signs are related by a global gauge transformation. 
	The states of the localized spins are specified by the angular components as ${\mathbf{S}_i}=(\sin\theta_i\cos\phi_i,\sin\theta_i\sin\phi_i,\cos\theta_i)$. The second part of the Hamiltonian~\eqref{equ:hamiltonian} represents the classical localized spin part:
	
	\begin{equation}
	\label{equ:clas-part}
	{\mathcal{\hat{H}}_c}= {\mathcal{\hat{H}}_{ex}} + {\mathcal{\hat{H}}_{DM}} + {\mathcal{\hat{H}}_H},
	\end{equation}	
   where ${\mathcal{\hat{H}}_{ex}}$ expresses the classical Heisenberg interaction between the localized spins,
	${\mathcal{\hat{H}}_{ex}}= \sum_{\avg{i,j}}J_{ij}\mathbf{S}_i \cdot \mathbf{S}_j.$
	${\mathcal{\hat{H}}_{DM}}$ describes the DM interaction,
	${\mathcal{\hat{H}}_{DM}}= \sum_{\avg{i,j}}\mathbf{D}_{ij}. (\mathbf{S}_i\times \mathbf{S}_j),$
	where $D_{ij}$ is the DM vector which is determined by the crystal symmetry of the lattice. 
	The precise values (directions and magnitude) of DM vectors will depend on the details of
	the crystal symmetry of each compound. In this study, we choose a generic set of DM vectors 
	and identify the minimal interactions that are necessary for stabilizing noncoplanar spin textures.
	The explicit form of the DM vectors on the different bonds is given in the caption of Fig.~\ref{fig:SSL-dm}. The last term in the Hamiltonian~\eqref{equ:clas-part} is the Zeeman term for the localized spins due to an external (longitudinal) magnetic field,
	${\mathcal{\hat{H}}_H}=-h^z\sum_i S_i^z $.
	A Zeeman term for the itinerant electrons is not included explicitly since the instantaneous spin orientation of the electrons is determined completely by the local moments in the large $J_K$ limit that we consider in this study.
	Hereafter, the parameters with primes represent the interactions on diagonal bonds, while the unprimed ones refer to the axial bonds. 
			
	\section{Method and observables}\label{sec:methods}
				
	To investigate the above model, we use an unbiased Monte Carlo (MC) method that has been used previously in the study of similar models~\cite{ishizuka-2012,yasu-2010,ishizuka-2013,ishizuka-2015}. A brief review of this method is presented here closely following Refs.~\onlinecite{motome-1999,furukawa-2004}.  The dynamics of large localized moments of the rare-earth ions is slow compared to itinerant electrons, and accordingly,  we can decouple their dynamics from that of the itinerant electrons. While studying the latter, we treat the local moments as static classical fields at each site. The electronic part of the Hamiltonian is bilinear in fermionic operators. Using the single-electron basis, ${\mathcal{\hat{H}}_e}$ can be represented as a $2\mathit{N}\times 2\mathit{N}$ matrix for a fixed configuration of classical localized spins, where $\mathit{N}$ is the number of sites.
					
	In order to explore the thermodynamic properties, we write the partition function for the whole system by taking two traces,
		
	\begin{multline}
	\label{equ:full trace}
	\mathit{Z}=\trace_{S}\trace_I \exp\{{-\beta[\mathcal{\hat{H}}_e}(\{x_r\})-\mu \hat{n}_e]\}\\
	\times \exp[-\beta ({\mathcal{\hat{H}}_c})],
	\end{multline}
	where $\trace_{S}$ and $\trace_I$ represent the traces over the classical localized spins  denoted by $\{x_r\}$ and the itinerant-electron degrees of freedom, respectively. The trace over itinerant-electron degrees of freedom can easily be calculated by the numerical diagonalization of Hamiltonian matrix ${\mathcal{\hat{H}}_e}$ for a fixed configuration of localized spins $\{x_r\}$,
		
	\begin{multline}
	\label{equ:trace}
	\trace_I \exp\{{-\beta[\mathcal{\hat{H}}_e}(\{x_r\})-\mu \hat{n}_e]\}\\
	\equiv\prod_\nu(1+\exp\{-\beta[\varepsilon_\nu(\{x_r\})-\mu]\}),
	\end{multline}
	where $\mu$ is the chemical potential, $\beta=1/k_BT$ is the inverse temperature, and $\hat{n}_e=\frac{1}{2N}\sum_{i\sigma}c_{i\sigma}^\dagger c_{i\sigma}$ is the number density operator for conduction electrons. The partition function for the whole system then takes the form
			
	\begin{equation}
	\label{equ:part func}
	\mathit{Z} =\trace_S \exp[-\mathit{S_{eff}}(\{x_r\})-\beta ({\mathcal{\hat{H}}_c})].
	\end{equation}
	The corresponding effective action is
	$\mathit{S_{eff}}(\{x_r\})=\sum_\nu\mathit{F}(y)$, where $\mathit{F}(y)=-\ln[1+\exp\{-\beta(y-\mu)\}]$. The grand-canonical trace over localized spin degrees of freedom is evaluated by sampling the spin configuration space using a MC method.  The probability distribution for a particular configuration of localized spins $\{x_r\}$ can be written as
	\begin{equation}
	\label{equ:prob dist}
	\mathit{P(\{x_r\})}\propto \exp[-\mathit{S_{eff}}(\{x_r\})-\beta( {\mathcal{\hat{H}}_c})].
	\end{equation}
			
	The thermodynamic quantities that depend on localized spins are calculated by the thermal averages of spin configurations, while the quantities that are associated with itinerant electrons are calculated from the eigenvalues and eigenfunctions of ${\mathcal{\hat{H}}_e}(\{x_r\})$. We start the simulations with a random configuration of localized spins $\{x_r\}$ and calculate Boltzmann action
	$\mathit{S_{eff}}(\{x_r\})$ for this configuration. The spin configuration is updated via the Metropolis algorithm based on the change in the effective actions of the configurations resulting from random updates, $\Delta \mathit{S_{eff}}=\mathit{S_{eff}}(\{x_r'\})-\mathit{S_{eff}}(\{x_r\})$. To identify magnetic orderings we calculate the magnetization per unit site as well as the spin structure factor, which is the Fourier transform of the spin-spin 
	correlation function,		
	\begin{equation}
	\label{equ:str fact}
	S(\mathbf{q})=\frac{1}{N}\sum_{i,j}\avg {\mathbf{S}_i \cdot \mathbf{S}_j} \exp [i\mathbf{q} \cdot \mathbf{r}_{ij}],
	\end{equation}
	where $\mathbf{r}_{ij}$ is the position vector from the $\mathit{i}$th to $\mathit{j}$th site and $\avg{\cdot}$ represents the thermal average over the grand-canonical ensemble. In order to describe the evolution of the ground state under the influence of an external magnetic field, we calculate the
	magnetization per site, defined as
			 
	\begin{equation}
	\label{equ:magnet02}
	m=\sqrt{\avg{\left (\frac{\sum_i \mathbf{S}_i}{N}\right )^2}}.
	\end{equation}
	To elucidate the difference between topological trivial and nontrivial states we evaluate the local scalar spin chirality. On a triangle the chirality is defined as
	\begin{equation}
	\label{equ:chirality}
	\chi_\bigtriangleup=\mathbf{S}_i \cdot (\mathbf{S}_{j} \times \mathbf{S}_{k}).
	\end{equation} 
	
%
%
%

	\noindent We use the total static spin chirality 
$\chi=\frac{1}{N_u}\sum_{\bigtriangleup} \chi_\bigtriangleup$ (where the sum is over all the 
triangles formed on the plaquettes with diagonal bonds and $N_u$ is the number of unit cells of SSL) as a quantitative measure of chiral order. This quantity is zero for collinear or coplanar magnetic states such as ferromagnetic (FM), antiferromagnetic (AFM), and pure flux states, whereas it is nonzero for {\em noncoplanar} configurations, e.g., all-out and  three-in--one-out states observed in pyrochlores. Finally, as an additional characterization of the chiral nature of the spin configurations, we measure the circulation of the in-plane components around each square plaquette as $f_m = \sum_\square \mathbf{S}_i\cdot \mathbf{r}_{ij} $, where $\mathbf{S}_i$ is the spin at site $i$ and $\mathbf{r}_{ij}$ are the vectors connecting sites $i$ and $j$ around the square plaquette in a counterclockwise direction. A nonzero circulation identifies a flux configuration of the local moments.

	\section{Results}\label{sec:results}
	
	Simulations of the Hamiltonian \eqref{equ:hamiltonian} are performed in lattices of dimension $L\times L$,
with $L=8 - 16$, over a wide range of parameters. For smaller lattices, we use the exact-diagonalization Monte Carlo (ED-MC) method where the full Hamiltonian is diagonalized to calculate the effective action for each MC step. For the larger lattices, we use 
the traveling cluster approximation (TCA) 
method~\cite{kumar-2006,anamitra-2015,majumdar-2006,kumar-2005}: a 
$6\times6$ cluster of SSL is used to calculate the effective action for one MC sweep.
Once the system is equilibrated,  the thermal averages are evaluated by diagonalizing 
the full Hamiltonian matrix. To avoid the severe freezing of the localized spins that happens at low temperature and to speed up the 
equilibration, we use simulated annealing. For this, we start the simulations at a relatively high temperature 
$T=0.1$ with a random localized spin configuration and run the system for equilibration and then 
use the final configuration at this temperature to perform the equilibration at $T=0.08$. We repeat this 
process with a step of temperature $\Delta T=0.02$, finally calculating the thermal averages of the 
observables at temperature $T=0.02$. For the lattice sizes
studied, the thermal gap to the lowest excitation is greater than the finite-size gap at $T=0.02$. 
In other words, $T=0.02$ is sufficiently small that ground-state estimates of the measured
observables can be reliably obtained.  Measurements are done for $50\:000$ MC steps after $60\:000$ 
steps discarded in total for thermalization.

	With its multidimensional parameter space, the 
	Hamiltonian~\eqref{equ:hamiltonian} is expected to support a rich array of ground-state phases over different ranges of the parameters. 
	In the present work, we restrict our attention to the magnetic behavior at an electronic filling 
factor $\avg{n_e} = 1/2$, for which the system is in an insulating state. The choice for the rest of 
the Hamiltonian parameters is guided by experimental observation in real materials. The
electronic hopping matrix elements along the axial bonds are chosen as $t=1.0$; this sets the 
energy scale for the problem. The diagonal hopping is set to $t'=1.2$, and the values of the 
exchange interactions along the axial and diagonal bonds are set at $J=0.1$ and $J'=0.12$. This 
choice is motivated by the experimental observation of approximately equal bond lengths in the 
rare-earth tetraboride family of compounds. In most materials of relevance to the present model, strong DM interactions exist. 
While the exact nature of DM interaction depends on the crystal symmetries, we have chosen a generic 
form of DM interaction for our study. Indeed, investigating the role of DM interaction in stabilizing noncoplanar 
spin configurations is a central goal of the present study. Finally, following the experimental observation in other frustrated metallic magnets such as pyrochlores, the strength of the Kondo coupling is chosen to be the strongest energy scale in the 
problem, $J_K=8.0$.

		
\subsection{Effect of DM interaction}
In the first part of the study, a systematic variation of the different components of the DM
vectors is performed to identify the minimal set of vectors necessary for noncoplanar
configurations of the local moments. We study the effects of the DM vectors normal to
the plane of the lattice $D_\perp$ and two in-plane components, $D_\parallel$ and 
$D'$ (Fig~.\ref{fig:SSL-dm}). It is found that while $D_\perp$ is essential for the emergence of noncollinearity,
$D^\prime$ and $D_\parallel$ stabilize the noncoplanarity of the ground state. 
	
	\subsubsection{Role of $D_\perp$}
	
	\begin{figure}[hbt]
		\centering
		\includegraphics[clip,trim=0cm 0cm 0cm 0cm,width=0.49\textwidth]{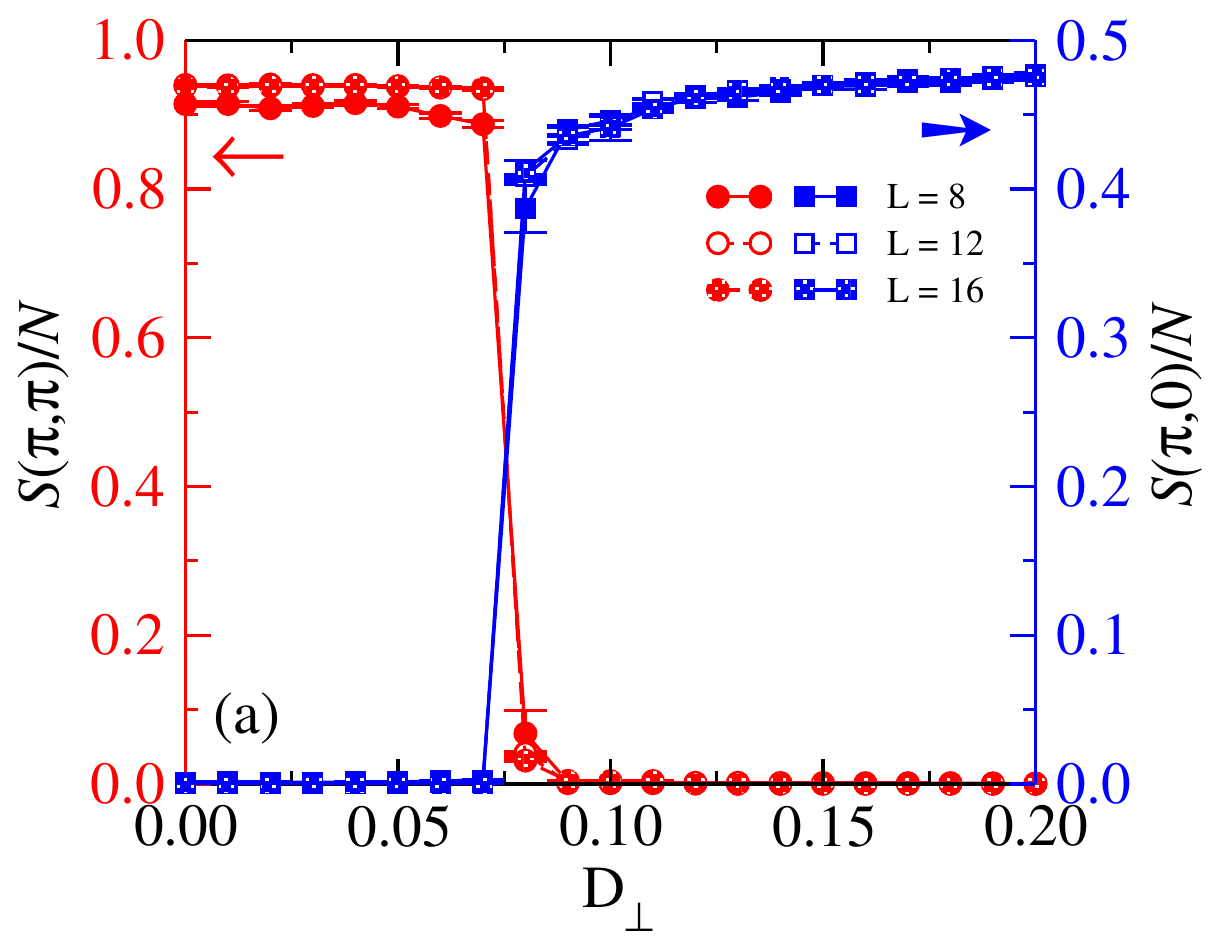}
		
		\includegraphics[clip,trim=0cm 0cm 0cm 0cm, width=0.49\textwidth]{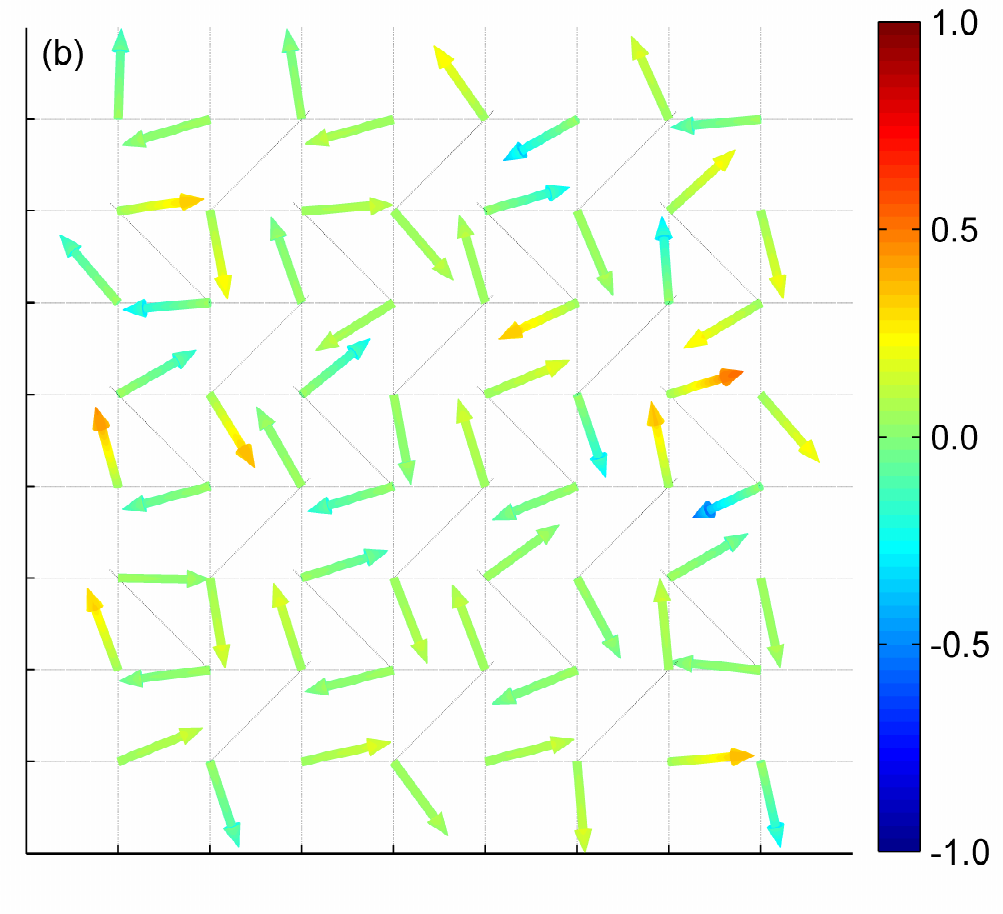}
		\caption{(Color online) (a) Evolution of the magnitude of the peaks in the static structure factor at $\mathbf{q}=(\pi,\pi)$ and $(\pi,0)$  as a function of $D_\perp$ for $8\times8$, $12\times12$, and $16\times16$ lattice sizes. (b) Snapshot of the real-space localized spin configuration for the $8\times8$ lattice at $D_\perp=0.11$, illustrating a flux state with a vanishingly small chirality. The results are obtained at $T=0.02$ while keeping $t=1.0$, $t^\prime=1.2$, $J=0.1$, $J^\prime=0.12$, $J_K=8.0$, $D_\parallel=0.0$, $D^\prime=0.0$, and $h^z=0.0$.}
		\label{fig:mag-chidper}
	\end{figure}
	
	We start our discussion by analyzing the nature of the magnetic ground state at zero field. The 
evolution of the ground state as $D_\perp$ is varied is shown in Fig.~\ref{fig:mag-chidper}(a), where we have plotted the magnitude of the peaks in the spin structure factor at $\mathbf{q}=(\pi,\pi)$ and $(\pi,0)$. At $D_\perp=0$, the ground state has predominantly longitudinal AFM order as the magnitude of the peak at $(\pi,\pi)$ is almost $1$ and the value of the magnetization is almost zero. With increasing the value of $D_\perp$, the ground state remains in the same phase up to a critical value $D_\perp^c \approx 0.07$, beyond which there is a discontinuous transition to a state characterized by a large magnitude of the peak at $(\pi,0)$. The static spin structure factor exhibits two sharp and equal-magnitude peaks at $(\pi,0)$ and $(0,\pi)$. Nominally, such features in the structure factor point towards a canted AFM state. However, the strong Kondo-like interaction precludes such ordering in the double-exchange model, as described in Ref.~\onlinecite{yamanaka-1998}.
The true nature of the ground state is illustrated by a snapshot of the real-space (periodic) equilibrium spin configuration obtained from the simulations and shown schematically in Fig.~\ref{fig:mag-chidper}(b). The in-plane components of the local moments are arranged in a near-ideal flux pattern along vanishingly small chirality; that is,  the magnetic ground state is a noncollinear flux state. The net magnetization and the chirality remain vanishingly small across the range of $D_\perp$ studied, further confirming the coplanar character of the flux state. Such complex spin textures are essential ingredients for the observation of unusual transport and electronic
phenomena as noncollinear and noncoplanar magnetic ordering of localized spins behave like emergent electromagnetic fields.

\begin{figure}[hbt]
	\centering
	\includegraphics[width=0.49\textwidth,trim=0cm 0cm 0cm 0cm,clip]{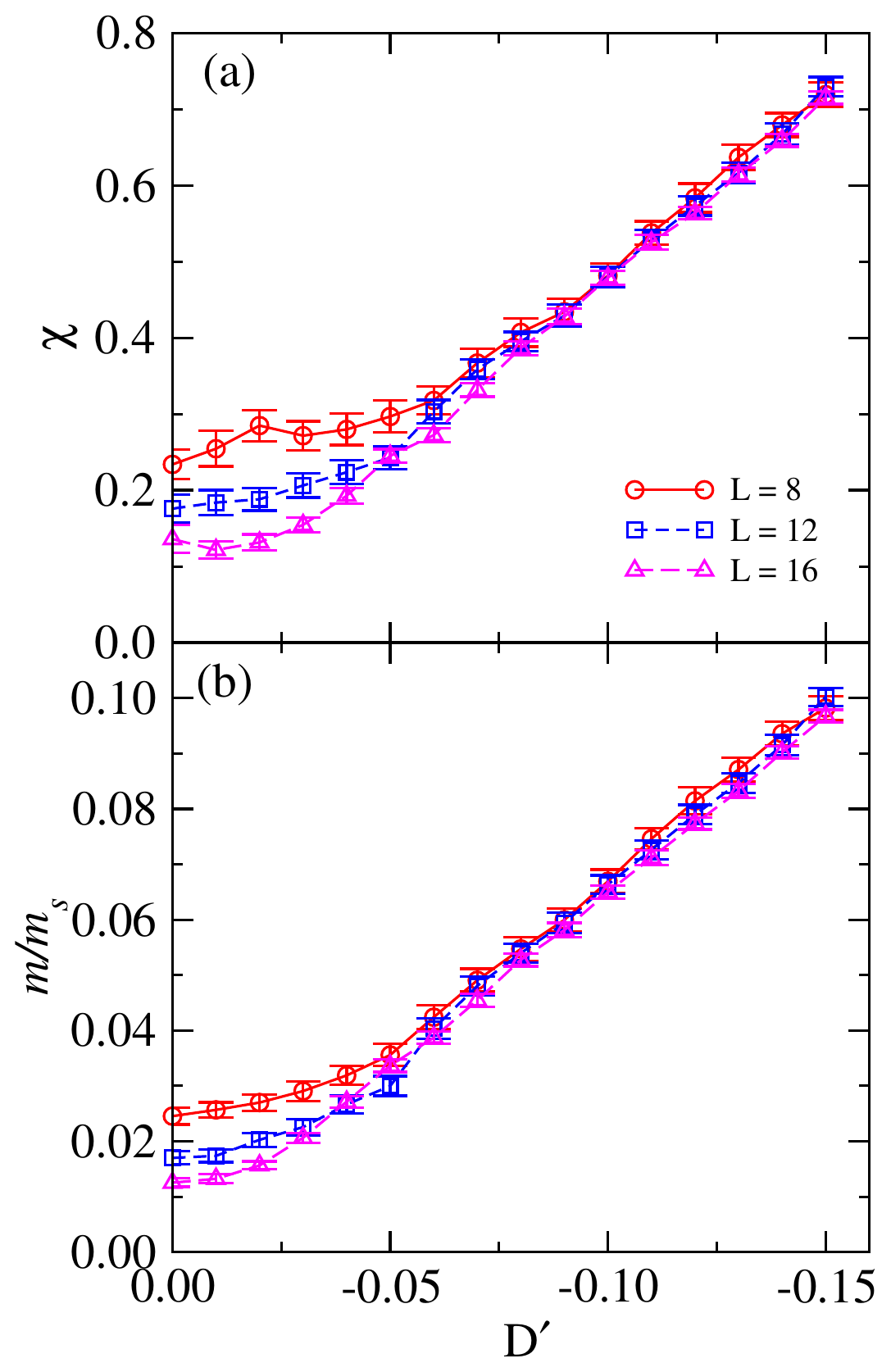}	
	\caption{(Color online) (a) Chirality per unit cell and (b) magnetization per unit site as a function of $D^\prime$ for $8\times8$, $12\times12$, and $16\times16$ lattice sizes. The MC results are obtained at $T=0.02$, $t=1.0$, $t^\prime=1.2$, $J=0.1$, $J^\prime=0.12$, $J_K=8.0$, $D_\parallel=0.0$, $D_\perp=0.11$, and $h^z=0.0$.}
	\label{fig:mag-chidpr}
\end{figure}
	
	\subsubsection{Role of $D'$}
	
	After finding the minimum value of $D_\perp$ that causes the phase transition to a 
noncollinear state, we discuss the effect of $D^\prime$ on stabilizing the noncoplanarity 
of the ground state. The results for chirality and magnetization per unit site as a function 
of $D^\prime$ are shown in Figs.~\ref{fig:mag-chidpr}(a) and \ref{fig:mag-chidpr}(b). For weak 
$D' (|D'| \lesssim 0.03)$, the chirality [Fig.~\ref{fig:mag-chidpr}(a)] decreases systematically with
increasing system size. This suggests that the chirality scales to zero in the thermodynamic
limit, but a careful finite-size scaling is needed  to ascertain that. For stronger $D'$,
the data are converged with system size and convincingly indicate a nonzero
chirality for the magnetic ground state. A strong value of $D^\prime$ results in an enlarged 
out-of-plane component of the localized spins making the ground state more canted.  
With increasing $D^\prime$ the noncoplanarity of the magnetic ordered state increases
monotonically [see Fig.~\ref{fig:mag-chidpr}(a)]. The same effect is observed in the 
behavior of the magnetization per unit site: $m/m_s$ decreases with increasing
system sizes at weak $D'$ but is finite (and nonzero) for $|D'| \gtrsim 0.03$ and
increases monotonically with $D^\prime$. The enlarged out-of-plane component of the
spins contributes to an increase in zero-field magnetization [shown in 
Fig.~\ref{fig:mag-chidpr}(b)]. The static spin structure factor  $S(\mathbf{q})$ at 
$D^\prime=-0.10$ (not shown here) exhibits a subdominant peak at $\mathbf{q}=(0,0)$, 
consistent with the finite net magnetization. Similarly, a real-space snapshot 
of the ground state (not shown here) shows that canting of spins increases with the 
introduction of $D^\prime$; we call this a canted flux state. It is worth mentioning that 
$D^\prime$ cannot induce a phase transition to a noncollinear flux state on its own; one 
always needs a nonzero value of $D_\perp$ for that. In contrast to pyrochlores in which the 
tetrahedral ordering of the local moments is fixed by the crystal-field effects, the canted flux state in our study arises dynamically from  competing interactions in the presence of 
geometric frustration. This enables us to control these complex magnetic orderings 
continually via an external magnetic field.

\begin{figure}[hbt]
	\centering
	\includegraphics[width=0.49\textwidth,trim=0cm 0cm 0cm 0cm,clip]{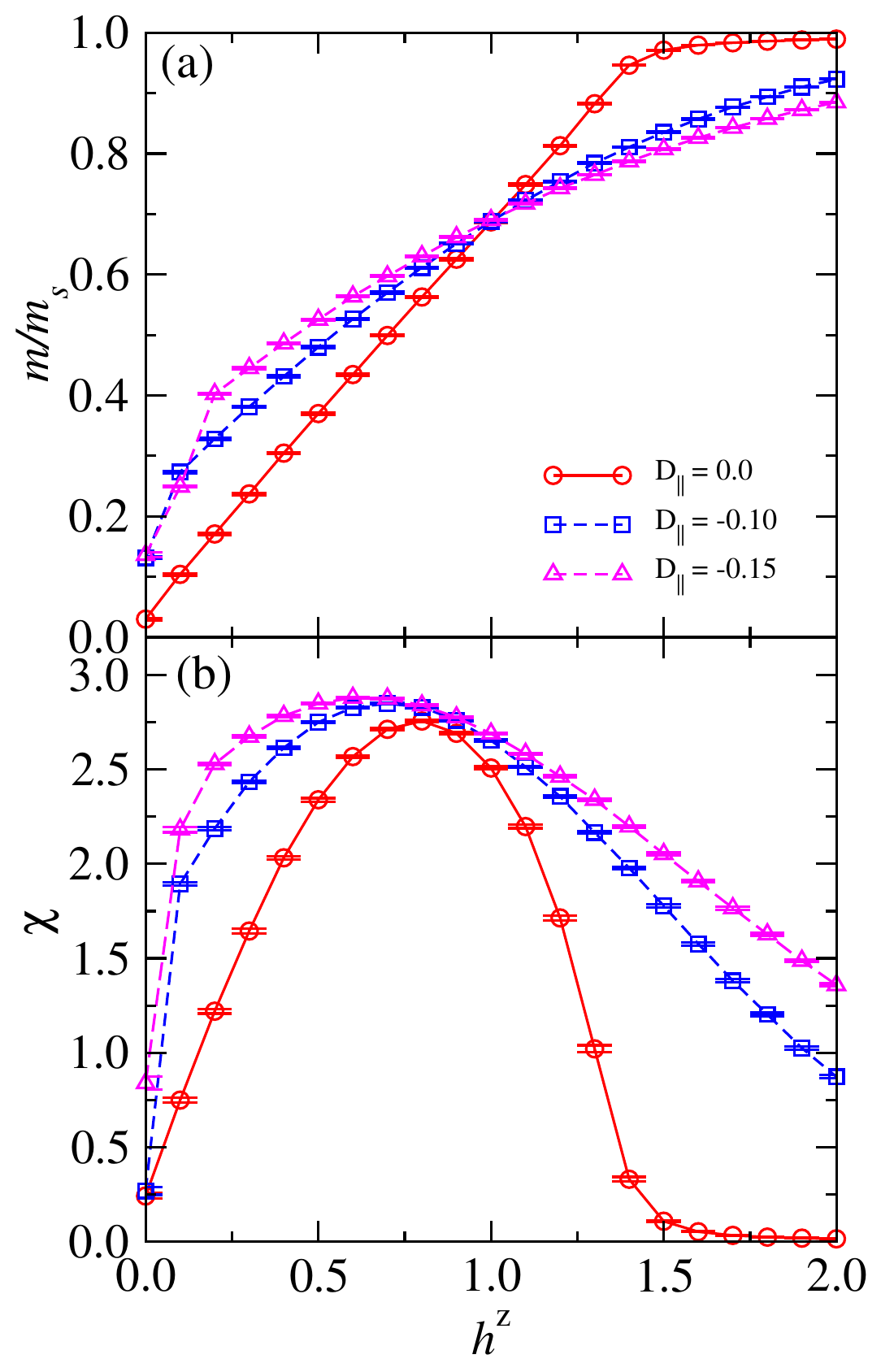}
	\caption{(Color online) (a) Magnetization per unit site and (b) chirality per unit cell as a function of external magnetic field for different values of $D_\parallel$. The results are obtained for the  $12\times12$ lattice at $T=0.02$, $t=1.0$, $t^\prime=1.2$, $J=0.1$, $J^\prime=0.12$, $J_K=8.0$, $D_\perp=0.11$, and $D^\prime=-0.05$.}
	\label{fig:mag-chipara}
\end{figure}

    \subsubsection{Role of $D_\parallel (h_z=0)$}
The other in-plane component of the DM vector $D_\parallel$
simply reinforces noncoplanar configurations driven by $D_\perp$ (increased $\chi$) 
and increases the uniform magnetization by enhancing the canting of the local 
moments out-of-$xy$ plane. 

    \subsection{Effects of an external magnetic field}
    
    \onecolumngrid
    
    \begin{figure}[hbt]
    	\centering
    	\includegraphics[clip,trim=0 0 0 0,width=\linewidth]{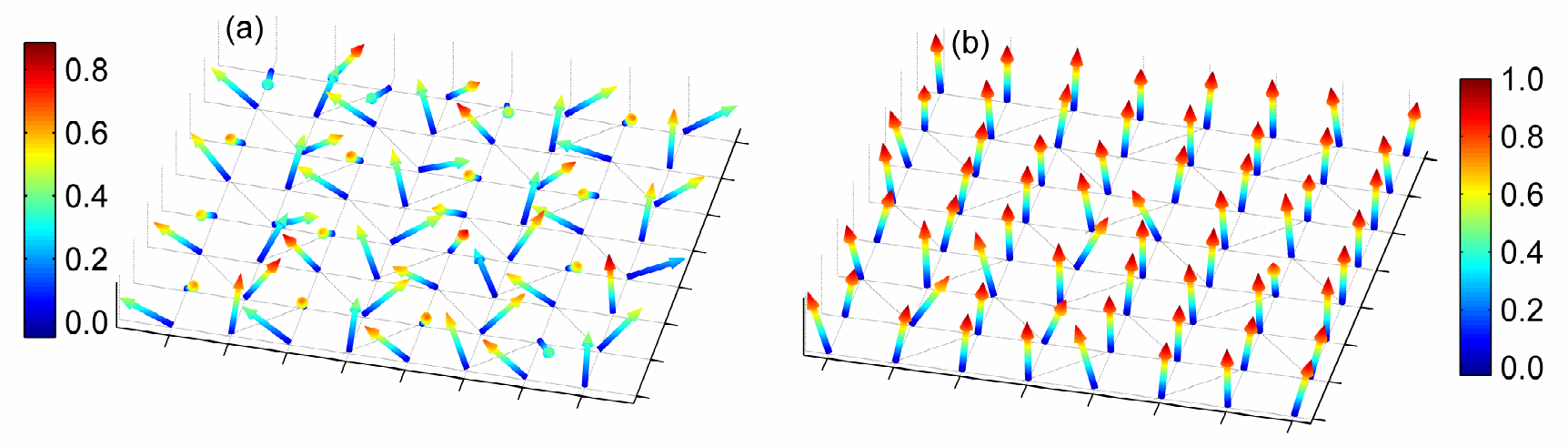}
    	\includegraphics[clip,trim=0 0 0 0,width=\linewidth]{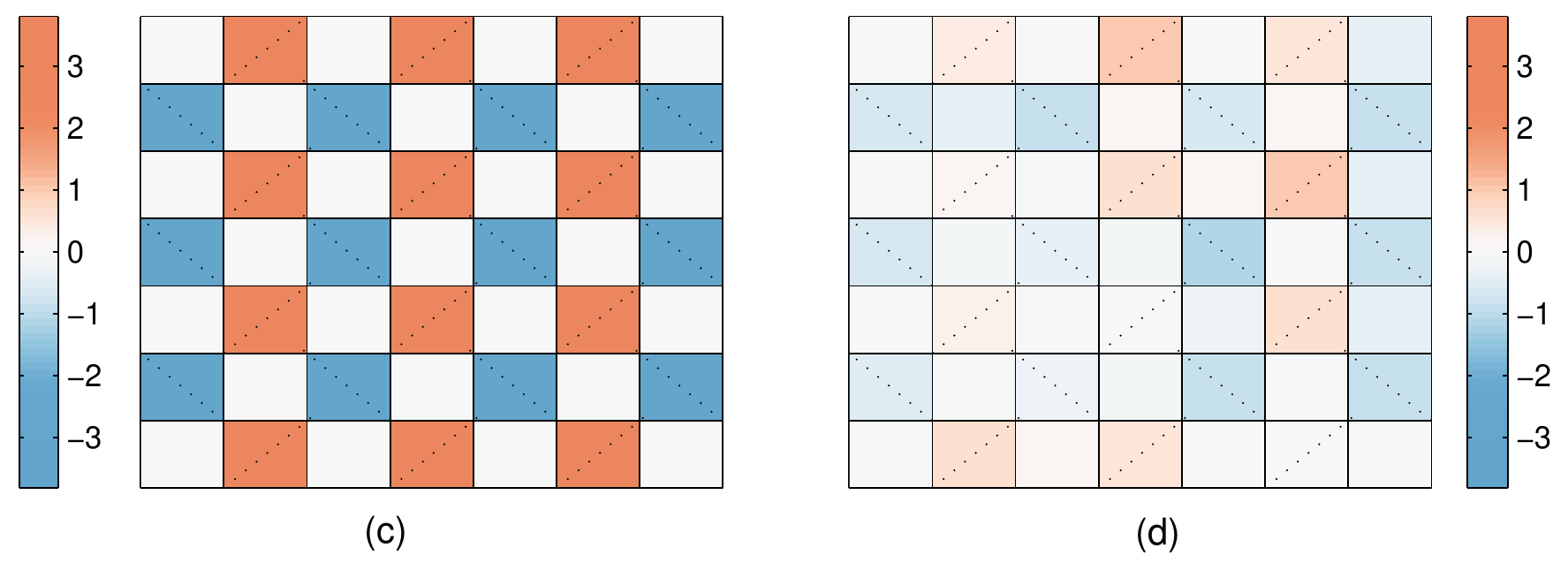}
    	\caption{(Color online) The real-space configurations of the localized spins for the $8\times8$ lattice shown at different values of magnetic field, (a) $h^z=0.8$ and (b) $h^z=1.6$, when local spins become almost polarized to the direction of magnetic field. The color bars beside the top plots indicate the out-of-plane component of the spin vector $S_i^z$. (c) and (d) Snapshots depicting the circulation of flux, clockwise or counterclockwise, on each plaquette for the same values of magnetic field. The MC simulations are performed at $T=0.02$, $t=1.0$, $t^\prime=1.2$, $J=0.1$, $J^\prime=0.12$, $J_K=8.0$, $D_\parallel=0.0$, $D_{\perp}=0.11$, and $D^\prime=-0.05$.}
    	\label{fig:confg-plot}
    \end{figure}
    \twocolumngrid 
    
	One of the most intriguing features of the canonical (purely magnetic) Shastry-Sutherland 
model is the appearance of magnetization plateaus in an applied magnetic field.  
DM interactions are expected to strongly modify the plateau structure. Our results show that 
magnetization plateaus are completely suppressed in the range of  simultaneous DM 
and Kondo-like interactions studied in this work. For $D_\perp \geq 0.08$ (which 
is necessary to stabilize the noncoplanar spin textures that we are interested in) and
in the absence of $D_\parallel$, the 
magnetization increases monotonically all the way to saturation. The ground state is
in a canted flux state at zero magnetic field. The peak at $(\pi,0)$ in the static 
structure factor, $S(\pi,0)/N$, decreases continuously to zero at a saturation
field strength of $h^z_s\approx 1.5$. With increasing field, the canting of the local
moments (which is nonzero but small at zero field due to the DM interaction) 
increases continuously until the local moments are fully polarized. Interestingly, the 
approach to saturation is distinct from that of a standard two-dimensional Heisenberg AFM in
the absence of an increased
slope in the magnetization curve, further underscoring the difference of the canted
flux state from a conventional canted AFM state. For nonzero $D_\parallel$, there is
a sharp increase in magnetization upon the application of a small longitudinal field 
($h^z \lesssim 0.1$) to a state with a finite $m/m_s$ that depends on the strength
of $D_\parallel$. Whether or not this happens via a discontinuous transition at 
$h^z =0$ is not clear from our results. Upon further increasing the longitudinal field,
the magnetization increases monotonically up to saturation, with the saturation field
increasing with increasing $D_\parallel$. The nature of the magnetic ground state 
remains a canted flux state throughout the field range. It remains to be seen if 
magnetization plateaus can be stabilized for any range of DM interaction and 
coupling between the local moments and the itinerant electrons.

	Finally, we discuss the topological nature of the magnetic ground state as it is 
tuned by an external field. Figure~\ref{fig:mag-chipara}(b) shows the evolution of the 
net static spin chirality with increasing magnetic field for a representative set of 
DM vectors where the zero-field ground state
is a canted flux state. With the increase 
of $D_\parallel$ the canting of the localized spins increases even at zero field. The 
ground-state spin texture remains noncoplanar in nature over the entire range of 
applied field strength up to saturation. The chirality increases monotonically up to 
an intermediate value of the applied field and then decreases continuously to zero at
saturation. The change in chirality is simply driven by the increasing canting of the spins along the direction of the applied field (and a subsequent decrease in the
magnitude of the in-plane component of the local moments). Once again, a snapshot
of local spin configurations elucidates the true nature of the magnetic ground state in
an applied field [Figs.~\ref{fig:confg-plot}(a) and \ref{fig:confg-plot}(b)]. The in-plane components of the local 
moments are arranged in a flux pattern on alternating plaquettes, whereas there
is a net canting of the longitudinal component parallel to the applied field. The 
transition to saturation is marked by the complete breaking of the flux pattern driven 
by the polarization of all the spins in the direction of magnetic field. 
The qualitative features of Figs.~\ref{fig:confg-plot}(a) and \ref{fig:confg-plot}(b) can be quantified partially in 
terms of the circulation of the in-plane components around each square plaquette, 
$f_m = \sum_\square \mathbf{S}_i\cdot \mathbf{r}_{ij} $. A nonzero circulation 
identifies a flux configuration of the local moments. Figures~\ref{fig:confg-plot}(c) and \ref{fig:confg-plot}(d) present
plots of the local circulation for the same sets of parameters as in Figs.~\ref{fig:confg-plot}(a) and \ref{fig:confg-plot}(b). For $h^z=0.8$ the circulation is equal in magnitude and opposite in sign for the plaquettes with diagonal bonds, whereas it is vanishingly small in the other plaquettes. In other words, the in-plane components form a flux pattern. The complete breaking of the flux pattern at $h_{c}\approx 1.6$ is reflected in the vanishingly small magnitude of the circulation not only for the plaquettes without diagonal bonds but also for the plaquettes with diagonal bonds [Fig.~\ref{fig:confg-plot}(d)], thus complementing the information inferred from Figs.~\ref{fig:confg-plot}(a) and \ref{fig:confg-plot}(b).

    \vspace{2.0em}
	\section{Summary}\label{sec:summary}
				
	To summarize, we have studied the Kondo lattice model with
additional DM interaction on the Shastry-Sutherland lattice. Our results show that complex, noncoplanar spin configurations can be generated dynamically from purely short range interactions 
and coupling to itinerant electrons. We conclude that DM interactions are necessary for the 
emergence of chiral spin configurations when the electronic spectrum is gapped; that is, the 
system is in an insulating state. We have carefully identified the minimal DM vectors necessary 
for the stabilization of noncoplanar configurations of the local moments. Furthermore, such noncoplanar structures can be tuned continually by applying an external magnetic field. These results provide insight into the origin and nature of topologically nontrivial magnetic  phases in metallic magnets. They will also be crucial in understanding the magnetic and electronic properties of the rare-earth tetraboride family of frustrated magnets. 
				
	\begin{acknowledgments}
	It is a pleasure to thank G. Alvarez and Y. Kato for useful discussions on the development of the numerical method. This work is partially supported by Grant No. MOE2014-T2-2-112 from the Ministry of Education, Singapore.
	\end{acknowledgments}
			
	\bibliographystyle{apsrev4-1}
	\bibliography{paperbib}

\begin{thebibliography}{40}%
\makeatletter
\providecommand \@ifxundefined [1]{%
 \@ifx{#1\undefined}
}%
\providecommand \@ifnum [1]{%
 \ifnum #1\expandafter \@firstoftwo
 \else \expandafter \@secondoftwo
 \fi
}%
\providecommand \@ifx [1]{%
 \ifx #1\expandafter \@firstoftwo
 \else \expandafter \@secondoftwo
 \fi
}%
\providecommand \natexlab [1]{#1}%
\providecommand \enquote  [1]{``#1''}%
\providecommand \bibnamefont  [1]{#1}%
\providecommand \bibfnamefont [1]{#1}%
\providecommand \citenamefont [1]{#1}%
\providecommand \href@noop [0]{\@secondoftwo}%
\providecommand \href [0]{\begingroup \@sanitize@url \@href}%
\providecommand \@href[1]{\@@startlink{#1}\@@href}%
\providecommand \@@href[1]{\endgroup#1\@@endlink}%
\providecommand \@sanitize@url [0]{\catcode `\\12\catcode `\$12\catcode
  `\&12\catcode `\#12\catcode `\^12\catcode `\_12\catcode `\%12\relax}%
\providecommand \@@startlink[1]{}%
\providecommand \@@endlink[0]{}%
\providecommand \url  [0]{\begingroup\@sanitize@url \@url }%
\providecommand \@url [1]{\endgroup\@href {#1}{\urlprefix }}%
\providecommand \urlprefix  [0]{URL }%
\providecommand \Eprint [0]{\href }%
\providecommand \doibase [0]{http://dx.doi.org/}%
\providecommand \selectlanguage [0]{\@gobble}%
\providecommand \bibinfo  [0]{\@secondoftwo}%
\providecommand \bibfield  [0]{\@secondoftwo}%
\providecommand \translation [1]{[#1]}%
\providecommand \BibitemOpen [0]{}%
\providecommand \bibitemStop [0]{}%
\providecommand \bibitemNoStop [0]{.\EOS\space}%
\providecommand \EOS [0]{\spacefactor3000\relax}%
\providecommand \BibitemShut  [1]{\csname bibitem#1\endcsname}%
\let\auto@bib@innerbib\@empty
\bibitem [{\citenamefont {Martin}\ and\ \citenamefont
  {Batista}(2008)}]{martin-2008}%
  \BibitemOpen
  \bibfield  {author} {\bibinfo {author} {\bibfnamefont {I.}~\bibnamefont
  {Martin}}\ and\ \bibinfo {author} {\bibfnamefont {C.~D.}\ \bibnamefont
  {Batista}},\ }\href {\doibase 10.1103/PhysRevLett.101.156402} {\bibfield
  {journal} {\bibinfo  {journal} {Phys. Rev. Lett.}\ }\textbf {\bibinfo
  {volume} {101}},\ \bibinfo {pages} {156402} (\bibinfo {year}
  {2008})}\BibitemShut {NoStop}%
\bibitem [{\citenamefont {Barros}\ \emph {et~al.}(2014)\citenamefont {Barros},
  \citenamefont {Venderbos}, \citenamefont {Chern},\ and\ \citenamefont
  {Batista}}]{barros-2014}%
  \BibitemOpen
  \bibfield  {author} {\bibinfo {author} {\bibfnamefont {K.}~\bibnamefont
  {Barros}}, \bibinfo {author} {\bibfnamefont {J.~W.~F.}\ \bibnamefont
  {Venderbos}}, \bibinfo {author} {\bibfnamefont {G.-W.}\ \bibnamefont
  {Chern}}, \ and\ \bibinfo {author} {\bibfnamefont {C.~D.}\ \bibnamefont
  {Batista}},\ }\href {\doibase 10.1103/PhysRevB.90.245119} {\bibfield
  {journal} {\bibinfo  {journal} {Phys. Rev. B}\ }\textbf {\bibinfo {volume}
  {90}},\ \bibinfo {pages} {245119} (\bibinfo {year} {2014})}\BibitemShut
  {NoStop}%
\bibitem [{\citenamefont {Ishizuka}\ and\ \citenamefont
  {Motome}(2013{\natexlab{a}})}]{ishizuka-2013ll}%
  \BibitemOpen
  \bibfield  {author} {\bibinfo {author} {\bibfnamefont {H.}~\bibnamefont
  {Ishizuka}}\ and\ \bibinfo {author} {\bibfnamefont {Y.}~\bibnamefont
  {Motome}},\ }\href {\doibase 10.1103/PhysRevB.88.081105} {\bibfield
  {journal} {\bibinfo  {journal} {Phys. Rev. B}\ }\textbf {\bibinfo {volume}
  {88}},\ \bibinfo {pages} {081105} (\bibinfo {year}
  {2013}{\natexlab{a}})}\BibitemShut {NoStop}%
\bibitem [{\citenamefont {Chern}(2015)}]{chern-2015}%
  \BibitemOpen
  \bibfield  {author} {\bibinfo {author} {\bibfnamefont {G.-W.}\ \bibnamefont
  {Chern}},\ }\href {\doibase 10.1142/S2010324715400068} {\bibfield  {journal}
  {\bibinfo  {journal} {SPIN}\ }\textbf {\bibinfo {volume} {05}},\ \bibinfo
  {pages} {1540006} (\bibinfo {year} {2015})}\BibitemShut {NoStop}%
\bibitem [{\citenamefont {Kato}\ \emph {et~al.}(2010)\citenamefont {Kato},
  \citenamefont {Martin},\ and\ \citenamefont {Batista}}]{yasu-2010}%
  \BibitemOpen
  \bibfield  {author} {\bibinfo {author} {\bibfnamefont {Y.}~\bibnamefont
  {Kato}}, \bibinfo {author} {\bibfnamefont {I.}~\bibnamefont {Martin}}, \ and\
  \bibinfo {author} {\bibfnamefont {C.~D.}\ \bibnamefont {Batista}},\ }\href
  {\doibase 10.1103/PhysRevLett.105.266405} {\bibfield  {journal} {\bibinfo
  {journal} {Phys. Rev. Lett.}\ }\textbf {\bibinfo {volume} {105}},\ \bibinfo
  {pages} {266405} (\bibinfo {year} {2010})}\BibitemShut {NoStop}%
\bibitem [{\citenamefont {Machida}\ \emph {et~al.}(2007)\citenamefont
  {Machida}, \citenamefont {Nakatsuji}, \citenamefont {Maeno}, \citenamefont
  {Tayama}, \citenamefont {Sakakibara},\ and\ \citenamefont
  {Onoda}}]{machida-2007}%
  \BibitemOpen
  \bibfield  {author} {\bibinfo {author} {\bibfnamefont {Y.}~\bibnamefont
  {Machida}}, \bibinfo {author} {\bibfnamefont {S.}~\bibnamefont {Nakatsuji}},
  \bibinfo {author} {\bibfnamefont {Y.}~\bibnamefont {Maeno}}, \bibinfo
  {author} {\bibfnamefont {T.}~\bibnamefont {Tayama}}, \bibinfo {author}
  {\bibfnamefont {T.}~\bibnamefont {Sakakibara}}, \ and\ \bibinfo {author}
  {\bibfnamefont {S.}~\bibnamefont {Onoda}},\ }\href {\doibase
  10.1103/PhysRevLett.98.057203} {\bibfield  {journal} {\bibinfo  {journal}
  {Phys. Rev. Lett.}\ }\textbf {\bibinfo {volume} {98}},\ \bibinfo {pages}
  {057203} (\bibinfo {year} {2007})}\BibitemShut {NoStop}%
\bibitem [{\citenamefont {Nakatsuji}\ \emph {et~al.}(2006)\citenamefont
  {Nakatsuji}, \citenamefont {Machida}, \citenamefont {Maeno}, \citenamefont
  {Tayama}, \citenamefont {Sakakibara}, \citenamefont {van Duijn},
  \citenamefont {Balicas}, \citenamefont {Millican}, \citenamefont {Macaluso},\
  and\ \citenamefont {Chan}}]{nakatsuji-2006}%
  \BibitemOpen
  \bibfield  {author} {\bibinfo {author} {\bibfnamefont {S.}~\bibnamefont
  {Nakatsuji}}, \bibinfo {author} {\bibfnamefont {Y.}~\bibnamefont {Machida}},
  \bibinfo {author} {\bibfnamefont {Y.}~\bibnamefont {Maeno}}, \bibinfo
  {author} {\bibfnamefont {T.}~\bibnamefont {Tayama}}, \bibinfo {author}
  {\bibfnamefont {T.}~\bibnamefont {Sakakibara}}, \bibinfo {author}
  {\bibfnamefont {J.}~\bibnamefont {van Duijn}}, \bibinfo {author}
  {\bibfnamefont {L.}~\bibnamefont {Balicas}}, \bibinfo {author} {\bibfnamefont
  {J.~N.}\ \bibnamefont {Millican}}, \bibinfo {author} {\bibfnamefont {R.~T.}\
  \bibnamefont {Macaluso}}, \ and\ \bibinfo {author} {\bibfnamefont {J.~Y.}\
  \bibnamefont {Chan}},\ }\href {\doibase 10.1103/PhysRevLett.96.087204}
  {\bibfield  {journal} {\bibinfo  {journal} {Phys. Rev. Lett.}\ }\textbf
  {\bibinfo {volume} {96}},\ \bibinfo {pages} {087204} (\bibinfo {year}
  {2006})}\BibitemShut {NoStop}%
\bibitem [{\citenamefont {Udagawa}\ and\ \citenamefont
  {Moessner}(2013)}]{udagawa-2013}%
  \BibitemOpen
  \bibfield  {author} {\bibinfo {author} {\bibfnamefont {M.}~\bibnamefont
  {Udagawa}}\ and\ \bibinfo {author} {\bibfnamefont {R.}~\bibnamefont
  {Moessner}},\ }\href {\doibase 10.1103/PhysRevLett.111.036602} {\bibfield
  {journal} {\bibinfo  {journal} {Phys. Rev. Lett.}\ }\textbf {\bibinfo
  {volume} {111}},\ \bibinfo {pages} {036602} (\bibinfo {year}
  {2013})}\BibitemShut {NoStop}%
\bibitem [{\citenamefont {Chern}(2010)}]{chern-2010}%
  \BibitemOpen
  \bibfield  {author} {\bibinfo {author} {\bibfnamefont {G.-W.}\ \bibnamefont
  {Chern}},\ }\href {\doibase 10.1103/PhysRevLett.105.226403} {\bibfield
  {journal} {\bibinfo  {journal} {Phys. Rev. Lett.}\ }\textbf {\bibinfo
  {volume} {105}},\ \bibinfo {pages} {226403} (\bibinfo {year}
  {2010})}\BibitemShut {NoStop}%
\bibitem [{\citenamefont {Akagi}\ and\ \citenamefont
  {Motome}(2011)}]{akagi-2011}%
  \BibitemOpen
  \bibfield  {author} {\bibinfo {author} {\bibfnamefont {Y.}~\bibnamefont
  {Akagi}}\ and\ \bibinfo {author} {\bibfnamefont {Y.}~\bibnamefont {Motome}},\
  }\href {http://stacks.iop.org/1742-6596/320/i=1/a=012059} {\bibfield
  {journal} {\bibinfo  {journal} {J. Phys. Conf. Ser.}\ }\textbf {\bibinfo
  {volume} {320}},\ \bibinfo {pages} {012059} (\bibinfo {year}
  {2011})}\BibitemShut {NoStop}%
\bibitem [{\citenamefont {Akagi}\ \emph {et~al.}(2012)\citenamefont {Akagi},
  \citenamefont {Udagawa},\ and\ \citenamefont {Motome}}]{akagi-2012}%
  \BibitemOpen
  \bibfield  {author} {\bibinfo {author} {\bibfnamefont {Y.}~\bibnamefont
  {Akagi}}, \bibinfo {author} {\bibfnamefont {M.}~\bibnamefont {Udagawa}}, \
  and\ \bibinfo {author} {\bibfnamefont {Y.}~\bibnamefont {Motome}},\ }\href
  {\doibase 10.1103/PhysRevLett.108.096401} {\bibfield  {journal} {\bibinfo
  {journal} {Phys. Rev. Lett.}\ }\textbf {\bibinfo {volume} {108}},\ \bibinfo
  {pages} {096401} (\bibinfo {year} {2012})}\BibitemShut {NoStop}%
\bibitem [{\citenamefont {Venderbos}\ \emph {et~al.}(2012)\citenamefont
  {Venderbos}, \citenamefont {Daghofer}, \citenamefont {van~den Brink},\ and\
  \citenamefont {Kumar}}]{venderbos-2012}%
  \BibitemOpen
  \bibfield  {author} {\bibinfo {author} {\bibfnamefont {J.~W.~F.}\
  \bibnamefont {Venderbos}}, \bibinfo {author} {\bibfnamefont {M.}~\bibnamefont
  {Daghofer}}, \bibinfo {author} {\bibfnamefont {J.}~\bibnamefont {van~den
  Brink}}, \ and\ \bibinfo {author} {\bibfnamefont {S.}~\bibnamefont {Kumar}},\
  }\href {\doibase 10.1103/PhysRevLett.109.166405} {\bibfield  {journal}
  {\bibinfo  {journal} {Phys. Rev. Lett.}\ }\textbf {\bibinfo {volume} {109}},\
  \bibinfo {pages} {166405} (\bibinfo {year} {2012})}\BibitemShut {NoStop}%
\bibitem [{\citenamefont {Karplus}\ and\ \citenamefont
  {Luttinger}(1954)}]{karplus-1954}%
  \BibitemOpen
  \bibfield  {author} {\bibinfo {author} {\bibfnamefont {R.}~\bibnamefont
  {Karplus}}\ and\ \bibinfo {author} {\bibfnamefont {J.~M.}\ \bibnamefont
  {Luttinger}},\ }\href {\doibase 10.1103/PhysRev.95.1154} {\bibfield
  {journal} {\bibinfo  {journal} {Phys. Rev.}\ }\textbf {\bibinfo {volume}
  {95}},\ \bibinfo {pages} {1154} (\bibinfo {year} {1954})}\BibitemShut
  {NoStop}%
\bibitem [{\citenamefont {Ye}\ \emph {et~al.}(1999)\citenamefont {Ye},
  \citenamefont {Kim}, \citenamefont {Millis}, \citenamefont {Shraiman},
  \citenamefont {Majumdar},\ and\ \citenamefont {Te\ifmmode \check{s}\else
  \v{s}\fi{}anovi\ifmmode~\acute{c}\else \'{c}\fi{}}}]{ye-1999}%
  \BibitemOpen
  \bibfield  {author} {\bibinfo {author} {\bibfnamefont {J.}~\bibnamefont
  {Ye}}, \bibinfo {author} {\bibfnamefont {Y.~B.}\ \bibnamefont {Kim}},
  \bibinfo {author} {\bibfnamefont {A.~J.}\ \bibnamefont {Millis}}, \bibinfo
  {author} {\bibfnamefont {B.~I.}\ \bibnamefont {Shraiman}}, \bibinfo {author}
  {\bibfnamefont {P.}~\bibnamefont {Majumdar}}, \ and\ \bibinfo {author}
  {\bibfnamefont {Z.}~\bibnamefont {Te\ifmmode \check{s}\else
  \v{s}\fi{}anovi\ifmmode~\acute{c}\else \'{c}\fi{}}},\ }\href {\doibase
  10.1103/PhysRevLett.83.3737} {\bibfield  {journal} {\bibinfo  {journal}
  {Phys. Rev. Lett.}\ }\textbf {\bibinfo {volume} {83}},\ \bibinfo {pages}
  {3737} (\bibinfo {year} {1999})}\BibitemShut {NoStop}%
\bibitem [{\citenamefont {Xiao}\ \emph {et~al.}(2010)\citenamefont {Xiao},
  \citenamefont {Chang},\ and\ \citenamefont {Niu}}]{niu-2010}%
  \BibitemOpen
  \bibfield  {author} {\bibinfo {author} {\bibfnamefont {D.}~\bibnamefont
  {Xiao}}, \bibinfo {author} {\bibfnamefont {M.-C.}\ \bibnamefont {Chang}}, \
  and\ \bibinfo {author} {\bibfnamefont {Q.}~\bibnamefont {Niu}},\ }\href
  {\doibase 10.1103/RevModPhys.82.1959} {\bibfield  {journal} {\bibinfo
  {journal} {Rev. Mod. Phys.}\ }\textbf {\bibinfo {volume} {82}},\ \bibinfo
  {pages} {1959} (\bibinfo {year} {2010})}\BibitemShut {NoStop}%
\bibitem [{\citenamefont {Taguchi}\ \emph {et~al.}(2001)\citenamefont
  {Taguchi}, \citenamefont {Oohara}, \citenamefont {Yoshizawa}, \citenamefont
  {Nagaosa},\ and\ \citenamefont {Tokura}}]{taguchi-2001}%
  \BibitemOpen
  \bibfield  {author} {\bibinfo {author} {\bibfnamefont {Y.}~\bibnamefont
  {Taguchi}}, \bibinfo {author} {\bibfnamefont {Y.}~\bibnamefont {Oohara}},
  \bibinfo {author} {\bibfnamefont {H.}~\bibnamefont {Yoshizawa}}, \bibinfo
  {author} {\bibfnamefont {N.}~\bibnamefont {Nagaosa}}, \ and\ \bibinfo
  {author} {\bibfnamefont {Y.}~\bibnamefont {Tokura}},\ }\href {\doibase
  10.1126/science.1058161} {\bibfield  {journal} {\bibinfo  {journal}
  {Science}\ }\textbf {\bibinfo {volume} {291}},\ \bibinfo {pages} {2573}
  (\bibinfo {year} {2001})}\BibitemShut {NoStop}%
\bibitem [{\citenamefont {Shastry}\ and\ \citenamefont
  {Sutherland}(1981)}]{shastry-1981}%
  \BibitemOpen
  \bibfield  {author} {\bibinfo {author} {\bibfnamefont {B.~S.}\ \bibnamefont
  {Shastry}}\ and\ \bibinfo {author} {\bibfnamefont {B.}~\bibnamefont
  {Sutherland}},\ }\href {\doibase
  http://dx.doi.org/10.1016/0378-4363(81)90838-X} {\bibfield  {journal}
  {\bibinfo  {journal} {Phys. B+C (Amsterdam, Neth.)}\ }\textbf {\bibinfo
  {volume} {108}},\ \bibinfo {pages} {1069 } (\bibinfo {year}
  {1981})}\BibitemShut {NoStop}%
\bibitem [{\citenamefont {Coleman}\ and\ \citenamefont
  {Nevidomskyy}(2010)}]{coleman-2010}%
  \BibitemOpen
  \bibfield  {author} {\bibinfo {author} {\bibfnamefont {P.}~\bibnamefont
  {Coleman}}\ and\ \bibinfo {author} {\bibfnamefont {A.~H.}\ \bibnamefont
  {Nevidomskyy}},\ }\href {\doibase 10.1007/s10909-010-0213-4} {\bibfield
  {journal} {\bibinfo  {journal} {J. Low Temp. Phys.}\ }\textbf {\bibinfo
  {volume} {161}},\ \bibinfo {pages} {182} (\bibinfo {year}
  {2010})}\BibitemShut {NoStop}%
\bibitem [{\citenamefont {Bernhard}\ \emph {et~al.}(2011)\citenamefont
  {Bernhard}, \citenamefont {Coqblin},\ and\ \citenamefont
  {Lacroix}}]{bernhard-2011}%
  \BibitemOpen
  \bibfield  {author} {\bibinfo {author} {\bibfnamefont {B.~H.}\ \bibnamefont
  {Bernhard}}, \bibinfo {author} {\bibfnamefont {B.}~\bibnamefont {Coqblin}}, \
  and\ \bibinfo {author} {\bibfnamefont {C.}~\bibnamefont {Lacroix}},\ }\href
  {\doibase 10.1103/PhysRevB.83.214427} {\bibfield  {journal} {\bibinfo
  {journal} {Phys. Rev. B}\ }\textbf {\bibinfo {volume} {83}},\ \bibinfo
  {pages} {214427} (\bibinfo {year} {2011})}\BibitemShut {NoStop}%
\bibitem [{\citenamefont {Pixley}\ \emph {et~al.}(2014)\citenamefont {Pixley},
  \citenamefont {Yu},\ and\ \citenamefont {Si}}]{pixley-2014}%
  \BibitemOpen
  \bibfield  {author} {\bibinfo {author} {\bibfnamefont {J.~H.}\ \bibnamefont
  {Pixley}}, \bibinfo {author} {\bibfnamefont {R.}~\bibnamefont {Yu}}, \ and\
  \bibinfo {author} {\bibfnamefont {Q.}~\bibnamefont {Si}},\ }\href {\doibase
  10.1103/PhysRevLett.113.176402} {\bibfield  {journal} {\bibinfo  {journal}
  {Phys. Rev. Lett.}\ }\textbf {\bibinfo {volume} {113}},\ \bibinfo {pages}
  {176402} (\bibinfo {year} {2014})}\BibitemShut {NoStop}%
\bibitem [{\citenamefont {Su}\ and\ \citenamefont {Sengupta}(2015)}]{lei-2015}%
  \BibitemOpen
  \bibfield  {author} {\bibinfo {author} {\bibfnamefont {L.}~\bibnamefont
  {Su}}\ and\ \bibinfo {author} {\bibfnamefont {P.}~\bibnamefont {Sengupta}},\
  }\href {\doibase 10.1103/PhysRevB.92.165431} {\bibfield  {journal} {\bibinfo
  {journal} {Phys. Rev. B}\ }\textbf {\bibinfo {volume} {92}},\ \bibinfo
  {pages} {165431} (\bibinfo {year} {2015})}\BibitemShut {NoStop}%
\bibitem [{\citenamefont {Siemensmeyer}\ \emph {et~al.}(2008)\citenamefont
  {Siemensmeyer}, \citenamefont {Wulf}, \citenamefont {Mikeska}, \citenamefont
  {Flachbart}, \citenamefont {Gab\'ani}, \citenamefont
  {Mat'a\ifmmode~\check{s}\else \v{s}\fi{}}, \citenamefont {Priputen},
  \citenamefont {Efdokimova},\ and\ \citenamefont
  {Shitsevalova}}]{siemensmeyer-2008}%
  \BibitemOpen
  \bibfield  {author} {\bibinfo {author} {\bibfnamefont {K.}~\bibnamefont
  {Siemensmeyer}}, \bibinfo {author} {\bibfnamefont {E.}~\bibnamefont {Wulf}},
  \bibinfo {author} {\bibfnamefont {H.-J.}\ \bibnamefont {Mikeska}}, \bibinfo
  {author} {\bibfnamefont {K.}~\bibnamefont {Flachbart}}, \bibinfo {author}
  {\bibfnamefont {S.}~\bibnamefont {Gab\'ani}}, \bibinfo {author}
  {\bibfnamefont {S.}~\bibnamefont {Mat'a\ifmmode~\check{s}\else \v{s}\fi{}}},
  \bibinfo {author} {\bibfnamefont {P.}~\bibnamefont {Priputen}}, \bibinfo
  {author} {\bibfnamefont {A.}~\bibnamefont {Efdokimova}}, \ and\ \bibinfo
  {author} {\bibfnamefont {N.}~\bibnamefont {Shitsevalova}},\ }\href {\doibase
  10.1103/PhysRevLett.101.177201} {\bibfield  {journal} {\bibinfo  {journal}
  {Phys. Rev. Lett.}\ }\textbf {\bibinfo {volume} {101}},\ \bibinfo {pages}
  {177201} (\bibinfo {year} {2008})}\BibitemShut {NoStop}%
\bibitem [{\citenamefont {Wierschem}\ \emph {et~al.}(2015)\citenamefont
  {Wierschem}, \citenamefont {Sunku}, \citenamefont {Kong}, \citenamefont
  {Ito}, \citenamefont {Canfield}, \citenamefont {Panagopoulos},\ and\
  \citenamefont {Sengupta}}]{keola-2015}%
  \BibitemOpen
  \bibfield  {author} {\bibinfo {author} {\bibfnamefont {K.}~\bibnamefont
  {Wierschem}}, \bibinfo {author} {\bibfnamefont {S.~S.}\ \bibnamefont
  {Sunku}}, \bibinfo {author} {\bibfnamefont {T.}~\bibnamefont {Kong}},
  \bibinfo {author} {\bibfnamefont {T.}~\bibnamefont {Ito}}, \bibinfo {author}
  {\bibfnamefont {P.~C.}\ \bibnamefont {Canfield}}, \bibinfo {author}
  {\bibfnamefont {C.}~\bibnamefont {Panagopoulos}}, \ and\ \bibinfo {author}
  {\bibfnamefont {P.}~\bibnamefont {Sengupta}},\ }\href {\doibase
  10.1103/PhysRevB.92.214433} {\bibfield  {journal} {\bibinfo  {journal} {Phys.
  Rev. B}\ }\textbf {\bibinfo {volume} {92}},\ \bibinfo {pages} {214433}
  (\bibinfo {year} {2015})}\BibitemShut {NoStop}%
\bibitem [{\citenamefont {Sunku}\ \emph {et~al.}(2016)\citenamefont {Sunku},
  \citenamefont {Kong}, \citenamefont {Ito}, \citenamefont {Canfield},
  \citenamefont {Shastry}, \citenamefont {Sengupta},\ and\ \citenamefont
  {Panagopoulos}}]{sai-2016}%
  \BibitemOpen
  \bibfield  {author} {\bibinfo {author} {\bibfnamefont {S.~S.}\ \bibnamefont
  {Sunku}}, \bibinfo {author} {\bibfnamefont {T.}~\bibnamefont {Kong}},
  \bibinfo {author} {\bibfnamefont {T.}~\bibnamefont {Ito}}, \bibinfo {author}
  {\bibfnamefont {P.~C.}\ \bibnamefont {Canfield}}, \bibinfo {author}
  {\bibfnamefont {B.~S.}\ \bibnamefont {Shastry}}, \bibinfo {author}
  {\bibfnamefont {P.}~\bibnamefont {Sengupta}}, \ and\ \bibinfo {author}
  {\bibfnamefont {C.}~\bibnamefont {Panagopoulos}},\ }\href {\doibase
  10.1103/PhysRevB.93.174408} {\bibfield  {journal} {\bibinfo  {journal} {Phys.
  Rev. B}\ }\textbf {\bibinfo {volume} {93}},\ \bibinfo {pages} {174408}
  (\bibinfo {year} {2016})}\BibitemShut {NoStop}%
\bibitem [{\citenamefont {Suzuki}\ \emph {et~al.}(2010)\citenamefont {Suzuki},
  \citenamefont {Tomita}, \citenamefont {Kawashima},\ and\ \citenamefont
  {Sengupta}}]{suzuki-2010}%
  \BibitemOpen
  \bibfield  {author} {\bibinfo {author} {\bibfnamefont {T.}~\bibnamefont
  {Suzuki}}, \bibinfo {author} {\bibfnamefont {Y.}~\bibnamefont {Tomita}},
  \bibinfo {author} {\bibfnamefont {N.}~\bibnamefont {Kawashima}}, \ and\
  \bibinfo {author} {\bibfnamefont {P.}~\bibnamefont {Sengupta}},\ }\href
  {\doibase 10.1103/PhysRevB.82.214404} {\bibfield  {journal} {\bibinfo
  {journal} {Phys. Rev. B}\ }\textbf {\bibinfo {volume} {82}},\ \bibinfo
  {pages} {214404} (\bibinfo {year} {2010})}\BibitemShut {NoStop}%
\bibitem [{\citenamefont {Mat'aš}\ \emph {et~al.}(2010)\citenamefont
  {Mat'aš}, \citenamefont {Siemensmeyer}, \citenamefont {Wheeler},
  \citenamefont {Wulf}, \citenamefont {Beyer}, \citenamefont {Hermannsdörfer},
  \citenamefont {Ignatchik}, \citenamefont {Uhlarz}, \citenamefont {Flachbart},
  \citenamefont {Gabáni}, \citenamefont {Priputen}, \citenamefont
  {Efdokimova},\ and\ \citenamefont {Shitsevalova}}]{matas-2010}%
  \BibitemOpen
  \bibfield  {author} {\bibinfo {author} {\bibfnamefont {S.}~\bibnamefont
  {Mat'aš}}, \bibinfo {author} {\bibfnamefont {K.}~\bibnamefont
  {Siemensmeyer}}, \bibinfo {author} {\bibfnamefont {E.}~\bibnamefont
  {Wheeler}}, \bibinfo {author} {\bibfnamefont {E.}~\bibnamefont {Wulf}},
  \bibinfo {author} {\bibfnamefont {R.}~\bibnamefont {Beyer}}, \bibinfo
  {author} {\bibfnamefont {T.}~\bibnamefont {Hermannsdörfer}}, \bibinfo
  {author} {\bibfnamefont {O.}~\bibnamefont {Ignatchik}}, \bibinfo {author}
  {\bibfnamefont {M.}~\bibnamefont {Uhlarz}}, \bibinfo {author} {\bibfnamefont
  {K.}~\bibnamefont {Flachbart}}, \bibinfo {author} {\bibfnamefont
  {S.}~\bibnamefont {Gabáni}}, \bibinfo {author} {\bibfnamefont
  {P.}~\bibnamefont {Priputen}}, \bibinfo {author} {\bibfnamefont
  {A.}~\bibnamefont {Efdokimova}}, \ and\ \bibinfo {author} {\bibfnamefont
  {N.}~\bibnamefont {Shitsevalova}},\ }\href
  {http://stacks.iop.org/1742-6596/200/i=3/a=032041} {\bibfield  {journal}
  {\bibinfo  {journal} {J. Phys. Conf. Ser.}\ }\textbf {\bibinfo {volume}
  {200}},\ \bibinfo {pages} {032041} (\bibinfo {year} {2010})}\BibitemShut
  {NoStop}%
\bibitem [{\citenamefont {Suzuki}\ \emph {et~al.}(2009)\citenamefont {Suzuki},
  \citenamefont {Tomita},\ and\ \citenamefont {Kawashima}}]{suzuki-2009}%
  \BibitemOpen
  \bibfield  {author} {\bibinfo {author} {\bibfnamefont {T.}~\bibnamefont
  {Suzuki}}, \bibinfo {author} {\bibfnamefont {Y.}~\bibnamefont {Tomita}}, \
  and\ \bibinfo {author} {\bibfnamefont {N.}~\bibnamefont {Kawashima}},\ }\href
  {\doibase 10.1103/PhysRevB.80.180405} {\bibfield  {journal} {\bibinfo
  {journal} {Phys. Rev. B}\ }\textbf {\bibinfo {volume} {80}},\ \bibinfo
  {pages} {180405} (\bibinfo {year} {2009})}\BibitemShut {NoStop}%
\bibitem [{\citenamefont {Iga}\ \emph {et~al.}(2007)\citenamefont {Iga},
  \citenamefont {Shigekawa}, \citenamefont {Hasegawa}, \citenamefont
  {Michimura}, \citenamefont {Takabatake}, \citenamefont {Yoshii},
  \citenamefont {Yamamoto}, \citenamefont {Hagiwara},\ and\ \citenamefont
  {Kindo}}]{iga-2007}%
  \BibitemOpen
  \bibfield  {author} {\bibinfo {author} {\bibfnamefont {F.}~\bibnamefont
  {Iga}}, \bibinfo {author} {\bibfnamefont {A.}~\bibnamefont {Shigekawa}},
  \bibinfo {author} {\bibfnamefont {Y.}~\bibnamefont {Hasegawa}}, \bibinfo
  {author} {\bibfnamefont {S.}~\bibnamefont {Michimura}}, \bibinfo {author}
  {\bibfnamefont {T.}~\bibnamefont {Takabatake}}, \bibinfo {author}
  {\bibfnamefont {S.}~\bibnamefont {Yoshii}}, \bibinfo {author} {\bibfnamefont
  {T.}~\bibnamefont {Yamamoto}}, \bibinfo {author} {\bibfnamefont
  {M.}~\bibnamefont {Hagiwara}}, \ and\ \bibinfo {author} {\bibfnamefont
  {K.}~\bibnamefont {Kindo}},\ }\href {\doibase
  http://dx.doi.org/10.1016/j.jmmm.2006.10.476} {\bibfield  {journal} {\bibinfo
   {journal} {J. Magn. Magn. Mater.}\ }\textbf {\bibinfo {volume} {310}},\
  \bibinfo {pages} {e443 } (\bibinfo {year} {2007})}\BibitemShut {NoStop}%
\bibitem [{\citenamefont {Michimura}\ \emph {et~al.}(2009)\citenamefont
  {Michimura}, \citenamefont {Shigekawa}, \citenamefont {Iga}, \citenamefont
  {Takabatake},\ and\ \citenamefont {Ohoyama}}]{shinji-2009}%
  \BibitemOpen
  \bibfield  {author} {\bibinfo {author} {\bibfnamefont {S.}~\bibnamefont
  {Michimura}}, \bibinfo {author} {\bibfnamefont {A.}~\bibnamefont
  {Shigekawa}}, \bibinfo {author} {\bibfnamefont {F.}~\bibnamefont {Iga}},
  \bibinfo {author} {\bibfnamefont {T.}~\bibnamefont {Takabatake}}, \ and\
  \bibinfo {author} {\bibfnamefont {K.}~\bibnamefont {Ohoyama}},\ }\href
  {\doibase 10.1143/JPSJ.78.024707} {\bibfield  {journal} {\bibinfo  {journal}
  {J. Phys. Soc. Jpn.}\ }\textbf {\bibinfo {volume} {78}},\ \bibinfo {pages}
  {024707} (\bibinfo {year} {2009})}\BibitemShut {NoStop}%
\bibitem [{\citenamefont {Shahzad}\ and\ \citenamefont
  {Sengupta}(2017)}]{shahzad-2017}%
  \BibitemOpen
  \bibfield  {author} {\bibinfo {author} {\bibfnamefont {M.}~\bibnamefont
  {Shahzad}}\ and\ \bibinfo {author} {\bibfnamefont {P.}~\bibnamefont
  {Sengupta}},\ }\href {\doibase 10.1103/PhysRevB.96.224402} {\bibfield
  {journal} {\bibinfo  {journal} {Phys. Rev. B}\ }\textbf {\bibinfo {volume}
  {96}},\ \bibinfo {pages} {224402} (\bibinfo {year} {2017})}\BibitemShut
  {NoStop}%
\bibitem [{\citenamefont {Ishizuka}\ and\ \citenamefont
  {Motome}(2012)}]{ishizuka-2012}%
  \BibitemOpen
  \bibfield  {author} {\bibinfo {author} {\bibfnamefont {H.}~\bibnamefont
  {Ishizuka}}\ and\ \bibinfo {author} {\bibfnamefont {Y.}~\bibnamefont
  {Motome}},\ }\href {\doibase 10.1103/PhysRevLett.108.257205} {\bibfield
  {journal} {\bibinfo  {journal} {Phys. Rev. Lett.}\ }\textbf {\bibinfo
  {volume} {108}},\ \bibinfo {pages} {257205} (\bibinfo {year}
  {2012})}\BibitemShut {NoStop}%
\bibitem [{\citenamefont {Ishizuka}\ and\ \citenamefont
  {Motome}(2013{\natexlab{b}})}]{ishizuka-2013}%
  \BibitemOpen
  \bibfield  {author} {\bibinfo {author} {\bibfnamefont {H.}~\bibnamefont
  {Ishizuka}}\ and\ \bibinfo {author} {\bibfnamefont {Y.}~\bibnamefont
  {Motome}},\ }\href {\doibase 10.1103/PhysRevB.87.081105} {\bibfield
  {journal} {\bibinfo  {journal} {Phys. Rev. B}\ }\textbf {\bibinfo {volume}
  {87}},\ \bibinfo {pages} {081105} (\bibinfo {year}
  {2013}{\natexlab{b}})}\BibitemShut {NoStop}%
\bibitem [{\citenamefont {Ishizuka}\ and\ \citenamefont
  {Motome}(2015)}]{ishizuka-2015}%
  \BibitemOpen
  \bibfield  {author} {\bibinfo {author} {\bibfnamefont {H.}~\bibnamefont
  {Ishizuka}}\ and\ \bibinfo {author} {\bibfnamefont {Y.}~\bibnamefont
  {Motome}},\ }\href {\doibase 10.1103/PhysRevB.91.085110} {\bibfield
  {journal} {\bibinfo  {journal} {Phys. Rev. B}\ }\textbf {\bibinfo {volume}
  {91}},\ \bibinfo {pages} {085110} (\bibinfo {year} {2015})}\BibitemShut
  {NoStop}%
\bibitem [{\citenamefont {Motome}\ and\ \citenamefont
  {Furukawa}(1999)}]{motome-1999}%
  \BibitemOpen
  \bibfield  {author} {\bibinfo {author} {\bibfnamefont {Y.}~\bibnamefont
  {Motome}}\ and\ \bibinfo {author} {\bibfnamefont {N.}~\bibnamefont
  {Furukawa}},\ }\href {\doibase 10.1143/JPSJ.68.3853} {\bibfield  {journal}
  {\bibinfo  {journal} {J. Phys. Soc. Jpn.}\ }\textbf {\bibinfo {volume}
  {68}},\ \bibinfo {pages} {3853} (\bibinfo {year} {1999})}\BibitemShut
  {NoStop}%
\bibitem [{\citenamefont {Furukawa}\ and\ \citenamefont
  {Motome}(2004)}]{furukawa-2004}%
  \BibitemOpen
  \bibfield  {author} {\bibinfo {author} {\bibfnamefont {N.}~\bibnamefont
  {Furukawa}}\ and\ \bibinfo {author} {\bibfnamefont {Y.}~\bibnamefont
  {Motome}},\ }\href {\doibase 10.1143/JPSJ.73.1482} {\bibfield  {journal}
  {\bibinfo  {journal} {J. Phys. Soc. Jpn.}\ }\textbf {\bibinfo {volume}
  {73}},\ \bibinfo {pages} {1482} (\bibinfo {year} {2004})}\BibitemShut
  {NoStop}%
\bibitem [{\citenamefont {Kumar}\ and\ \citenamefont
  {Majumdar}(2006{\natexlab{a}})}]{kumar-2006}%
  \BibitemOpen
  \bibfield  {author} {\bibinfo {author} {\bibfnamefont {S.}~\bibnamefont
  {Kumar}}\ and\ \bibinfo {author} {\bibfnamefont {P.}~\bibnamefont
  {Majumdar}},\ }\href {\doibase 10.1140/epjb/e2006-00173-2} {\bibfield
  {journal} {\bibinfo  {journal} {Eur. Phys. J. B}\ }\textbf {\bibinfo {volume}
  {50}},\ \bibinfo {pages} {571} (\bibinfo {year}
  {2006}{\natexlab{a}})}\BibitemShut {NoStop}%
\bibitem [{\citenamefont {Mukherjee}\ \emph {et~al.}(2015)\citenamefont
  {Mukherjee}, \citenamefont {Patel}, \citenamefont {Bishop},\ and\
  \citenamefont {Dagotto}}]{anamitra-2015}%
  \BibitemOpen
  \bibfield  {author} {\bibinfo {author} {\bibfnamefont {A.}~\bibnamefont
  {Mukherjee}}, \bibinfo {author} {\bibfnamefont {N.~D.}\ \bibnamefont
  {Patel}}, \bibinfo {author} {\bibfnamefont {C.}~\bibnamefont {Bishop}}, \
  and\ \bibinfo {author} {\bibfnamefont {E.}~\bibnamefont {Dagotto}},\ }\href
  {\doibase 10.1103/PhysRevE.91.063303} {\bibfield  {journal} {\bibinfo
  {journal} {Phys. Rev. E}\ }\textbf {\bibinfo {volume} {91}},\ \bibinfo
  {pages} {063303} (\bibinfo {year} {2015})}\BibitemShut {NoStop}%
\bibitem [{\citenamefont {Kumar}\ and\ \citenamefont
  {Majumdar}(2006{\natexlab{b}})}]{majumdar-2006}%
  \BibitemOpen
  \bibfield  {author} {\bibinfo {author} {\bibfnamefont {S.}~\bibnamefont
  {Kumar}}\ and\ \bibinfo {author} {\bibfnamefont {P.}~\bibnamefont
  {Majumdar}},\ }\href {\doibase 10.1103/PhysRevLett.96.016602} {\bibfield
  {journal} {\bibinfo  {journal} {Phys. Rev. Lett.}\ }\textbf {\bibinfo
  {volume} {96}},\ \bibinfo {pages} {016602} (\bibinfo {year}
  {2006}{\natexlab{b}})}\BibitemShut {NoStop}%
\bibitem [{\citenamefont {Kumar}\ and\ \citenamefont
  {Majumdar}(2005)}]{kumar-2005}%
  \BibitemOpen
  \bibfield  {author} {\bibinfo {author} {\bibfnamefont {S.}~\bibnamefont
  {Kumar}}\ and\ \bibinfo {author} {\bibfnamefont {P.}~\bibnamefont
  {Majumdar}},\ }\href {\doibase 10.1103/PhysRevLett.94.136601} {\bibfield
  {journal} {\bibinfo  {journal} {Phys. Rev. Lett.}\ }\textbf {\bibinfo
  {volume} {94}},\ \bibinfo {pages} {136601} (\bibinfo {year}
  {2005})}\BibitemShut {NoStop}%
\bibitem [{\citenamefont {Yamanaka}\ \emph {et~al.}(1998)\citenamefont
  {Yamanaka}, \citenamefont {Koshibae},\ and\ \citenamefont
  {Maekawa}}]{yamanaka-1998}%
  \BibitemOpen
  \bibfield  {author} {\bibinfo {author} {\bibfnamefont {M.}~\bibnamefont
  {Yamanaka}}, \bibinfo {author} {\bibfnamefont {W.}~\bibnamefont {Koshibae}},
  \ and\ \bibinfo {author} {\bibfnamefont {S.}~\bibnamefont {Maekawa}},\ }\href
  {\doibase 10.1103/PhysRevLett.81.5604} {\bibfield  {journal} {\bibinfo
  {journal} {Phys. Rev. Lett.}\ }\textbf {\bibinfo {volume} {81}},\ \bibinfo
  {pages} {5604} (\bibinfo {year} {1998})}\BibitemShut {NoStop}%
\end{thebibliography}%
			
	\end{document}